\documentclass{article}

\usepackage{microtype}
\usepackage{graphicx}
\usepackage{subcaption}
\usepackage{booktabs} %
\usepackage{aas_macros}

\usepackage{hyperref}

\usepackage{amsmath}
\usepackage{amssymb}
\usepackage{mathtools}
\usepackage{amsthm}

\usepackage[capitalize,noabbrev]{cleveref}

\theoremstyle{plain}

\theoremstyle{definition}

\theoremstyle{remark}

\usepackage[accepted]{icml2024}

\icmltitlerunning{Predicting dark matter halo masses from simulated galaxy images and environments}

\begin{document}

\twocolumn[
\icmltitle{Predicting dark matter halo masses from \\simulated galaxy images and environments}

\icmlsetsymbol{equal}{*}

\begin{icmlauthorlist}
\icmlauthor{Austin J Larson}{jhucs}
\icmlauthor{John F Wu}{stsci,jhupa} 
\icmlauthor{Craig Jones}{jhucs} 
\end{icmlauthorlist}

\icmlaffiliation{jhucs}{Department of Computer Science, Johns Hopkins University, Baltimore, MD 21218, USA}
\icmlaffiliation{stsci}{Space Telescope Science Institute, 3700 San Martin Drive, Baltimore, MD 21218 USA}
\icmlaffiliation{jhupa}{Center for Astrophysical Sciences, Johns Hopkins University, Baltimore, MD 21218, USA}

\icmlcorrespondingauthor{John F. Wu}{jfwu@stsci.edu}

\icmlkeywords{Machine Learning, ICML}

\vskip 0.3in
]

\printAffiliationsAndNotice{} %

\begin{abstract}
Galaxies are theorized to form and co-evolve with their dark matter halos, such that their stellar masses and halo masses should be well-correlated. However, it is not known whether other observable galaxy features, such as their morphologies or large-scale environments, can be used to tighten the correlation between galaxy properties and halo masses. In this work, we train a baseline random forest model to predict halo mass using galaxy features from the Illustris TNG50 hydrodynamical simulation, and compare with convolutional neural networks (CNNs) and graph neural networks (GNNs) trained respectively using galaxy image cutouts and galaxy point clouds. The best baseline model has a root mean squared error (RMSE) of $0.310$ and mean absolute error (MAE) of $0.220$, compared to the CNN ($\text{RSME}=0.359$, $\text{MAE}=0.238$), GNN ($\text{RMSE}=0.248$, $\text{MAE}=0.158$), and a novel combined CNN+GNN ($\text{RMSE}=0.248$, $\text{MAE}=0.144$). The CNN is likely limited by our small data set, and we anticipate that the CNN and CNN+GNN would benefit from training on larger cosmological simulations. We conclude that deep learning models can leverage information from galaxy appearances and environment, beyond commonly used summary statistics, in order to better predict the halo mass.
\end{abstract}

\section{Introduction}

Galaxies are theorized to form inside and co-evolve with dark matter halos \citep{2018ARA&A..56..435W}.  The close relationship between galaxies and their halos has led to a tight relationship between galaxy stellar mass ($M_\star$) and dark matter halo mass ($M_{halo}$), which is known as the stellar mass--halo mass relation (SMHMR). The SMHMR can be calibrated by galaxy and halo properties derived from cosmological hydrodynamic simulations, or from other approaches such as semi-analytic models or empirical models \citep[e.g.,][]{2015ARA&A..53...51S,2019MNRAS.488.3143B}.

While galaxies are observable in the real Universe, dark matter properties are often determined through indirect measurements. For example, dark matter halo mass distributions can be inferred from the galaxy and galaxy cluster kinematics \citep[e.g.,][]{1933AcHPh...6..110Z,1980ApJ...238..471R} or through rare instances of gravitational lensing \citep[e.g.,][]{2006ApJ...648L.109C}. However, these constraints rely on detailed observations that are not widely available; in those cases, $M_{halo}$ can only be assumed by using the SMHMR to assign a halo mass given a galaxy stellar mass. 

However, it is likely that $M_{halo}$ depends on galaxy properties other than $M_\star$. We present an exploration of how $M_{halo}$ might be predicted from not just the stellar mass, but also galaxy morphology and large-scale environment. Using a galaxy sample from the Illustris TNG50-1 cosmological simulation \citep[hereafter TNG50;][]{2019ComAC...6....2N,2019MNRAS.490.3234N,2019MNRAS.490.3196P}, we test whether galaxy morphology and large-scale environment contains information that can improve (lower) the scatter in the SMHMR. 

A machine learning (ML) model such as a random forest can easily learn to predict $M_{halo}$ from $M_\star$ \citep[e.g.,][]{2016MNRAS.457.1162K}. We gauge the level of improvement in predicting $M_{halo}$ after we include galaxy morphological parameters (i.e., commonly used summary statistics) and/or environmental overdensity as features in ML models. We train deep neural networks to learn the morphological information directly from synthetic galaxy imaging, and to learn the environmental information directly from galaxy point clouds. In all these experiments, we rely on the simulated galaxy sample from TNG50 to calibrate $M_{halo}$ against other galaxy properties.

\subsection{Some notes on nomenclature}
Before we proceed, we define some terminology used in this paper. The terms \textit{galaxies} and \textit{halos} are used interchangeably, since every TNG50 galaxy studied here resides in a dark matter halo. Halos can be \textit{satellites} (also known as \textit{subhalos}) of a more massive halo, or they can be \textit{centrals}. The latter means that they gravitationally dominate their surroundings. 

We test several ML models that take galaxy \textit{features} as inputs. All features are derived from the TNG50 data (see Section~\ref{sec:data}). For the baseline models (described in Section~\ref{sec:baseline-models}), we employ morphological \textit{summary statistics}, i.e., scalar features commonly computed by astronomers from galaxy images that describe the appearances of galaxies. Environmental overdensity can also be described with summary statistics. Critically, these summary statistics are lossy descriptions of the underlying data, i.e., morphological summary statistics do not fully describe the information in galaxy images, and the overdensity summary statistic does not fully describe the information in galaxy environments. 

We compare baseline models against more sophisticated neural network models. To quantify these comparisons, we introduce several \textit{evaluation metrics} (see Section~\ref{sec:metrics}). Some metrics penalize large errors (i.e., \textit{outliers}) more severely, while other metrics are insensitive to outliers. For most metrics presented here (with the exception of the $R^2$ correlation coefficient), lower values are better. We optimize all models using the mean squared error (MSE) loss function, which also serves as a metric for comparison.

\subsection{Our contributions}

We present two findings that represent novel contributions to the field of astronomical machine learning:
\begin{itemize}
\item Simple parameterizations of galaxy morphology and environmental overdensity are informative for predicting galaxy halo masses. Combining these two sets of information using a joint CNN+GNN yields even better predictions.
\item Deep neural networks can learn additional information not contained in the simple features. Although our convolutional neural network requires a larger training sample to optimally and flexibly learn all the information within synthetic galaxy images, we find that our graph neural network dramatically lowers the $M_{halo}$ prediction error by learning environmental relationships from the galaxy point cloud.
\end{itemize}

\section{TNG50 Simulated Data} \label{sec:data}

\subsection{Galaxy catalogs}

We used simulated galaxy data from TNG50, the highest resolution hydrodynamical simulation in the IllustrisTNG Project. The high spatial resolution of TNG50 is necessary for adequately resolving galaxy morphologies and internal structures, but as a result, the simulation volume is small by cosmological standards (a box with $\sim 50$ Mpc per side). \citet{2019MNRAS.490.3196P} provide an overview of results from TNG50, and \citet{2021MNRAS.500.3957E} characterize the SMHMR in the TNG simulations in detail.

We downloaded subhalo catalogs from snapshot 99 ($z=0$), and we include both central and satellite galaxies in our sample.
Every dark matter halo in our sample contains a galaxy.
$M_{halo}$ is defined as the total mass of gravitationally bound dark matter particles, and $M_\star$ is defined as the total mass of gravitationally bound star particles in the simulation.
After administering quality flag cuts, we select galaxies with stellar mass $\log\left(M_{\star}/M_\odot\right) > 9.5$.
Our parent sample comprises 1,666 galaxies in the TNG50 volume.

\subsection{Image cutouts}

\citet{2019MNRAS.483.4140R} generate $gri$-band synthetic image cutouts for all galaxies with  $\log\left(M_{halo}/M_\odot\right) > 9$ by using \textsc{SKIRT} radiative transfer code \citep{2011ApJS..196...22B,2015A&C.....9...20C} and \citet{2003MNRAS.344.1000B} stellar population synthesis models designed to match observations from the Pan-STARRS 3$\pi$ Steradian Survey \citep{2016arXiv161205560C}. 

We crop or zero-pad the image cutouts to ensure that they are all the same size $3 \times 224 \times 224$, which is a common image size used in machine learning \citep{alexnet}. Each image must be further processed to add realistic observational effects. Following recommendations from \citet{2019MNRAS.483.4140R}, we convolve the images with an azimuthally symmetric Gaussian profile to match the Pan-STARRS survey imaging. We adopt a point spread function with 1.11, 1.21, and 1.31 arcsec full width at half-maximum (FWHM) in \textit{i}, \textit{r}, and \textit{g} bands, respectively.

\subsection{Morphological features} \label{sec:morphological-features}
In addition to generating synthetic galaxy images, \citet{2019MNRAS.483.4140R} provide morphological summary statistics derived from the image cutouts using the \texttt{statmorph} library. These morphological features include the radii at 20\%, 50\%, and 80\% of the galaxy light in circular apertures ($r_{20}$, $r_{50}$, $r_{80}$); Petrosian, Sersic, and half-light radii using best-fit elliptical apertures ($r_{Petro}$, $r_{Sersic}$, $r_{half}$); Sersic index ($n_{Sersic}$); Gini and $M_{20}$ statistics \citep{2004AJ....128..163L}; concentration, asymmetry, and smoothness (CAS) statistics \citep{Conselice2003}; and multimode, intensity, and deviation (MID) statistics \citep{2013MNRAS.434..282F}. We refer the reader to the original works for details on these morphological measurements \citep[see also Section 4 of][]{2019MNRAS.483.4140R}. We remove 117 galaxies that have unreliable morphological features based on a flag in the catalog, leaving 1549 galaxies in our sample. %

\subsection{Galaxy overdensity}

We compute a summary statistic for environmental overdensity, $\Delta_G$, by counting the number of galaxies within some $R_{max}$:
\begin{equation}
    \Delta_G = \left|\left\{H: d\left(G,H\right) < R_{max}\right\}\right|
\end{equation}
We choose $R_{max} = 3 \text{ Mpc}$, a scale that is most informative for describing the large-scale environment for the IllustrisTNG galaxy-halo connection \citep{wu2024galaxyhalo}. After we apply our selection criteria, the average galaxy density in the TNG50 catalogs is $2.8\times10^{-3}\ {\rm Mpc}^{-3}$.

\section{Experimental Design} \label{sec:experiments}

We aim to estimate the halo mass using progressively more complex techniques. For each experiment, we split the data into the same train, validation, and test split (Section~\ref{sec:split}). We minimize the validation loss in order to select optimization hyperparameters. After we choose these hyperparameters and train our models, we ``unblind'' the test set and report the results. In Section~\ref{sec:metrics}, we define our evaluation metrics. 

In Section~\ref{sec:baseline-models}, we describe the procedure for training baseline models to predict $M_{halo}$ using galaxy properties such as $M_\star$, morphological features, and overdensity parameter. These morphological and overdensity baseline models inform us whether there is extra information (beyond the simple baseline model) that is helpful for predicting halo mass. 

We further test whether morphological information can be learned empirically using CNNs (Section~\ref{sec:cnn}), and whether galaxy environmental information can be learned empirically using GNNs (Section~\ref{sec:gnn}). Finally, we present a novel combined CNN and GNN model that predicts $M_{halo}$ from the pixel-level image cutouts and galaxy point cloud (Section~\ref{sec:cnn+gnn}).

\subsection{Data Split} \label{sec:split}
\begin{figure*}[t]
\subcaptionbox{Each split and excluded galaxies shown in a $2d$ projected point cloud. The marker size is proportional to $M_{halo}$.\label{fig:point-cloud}}{
    \centering
    \includegraphics[width=80mm]{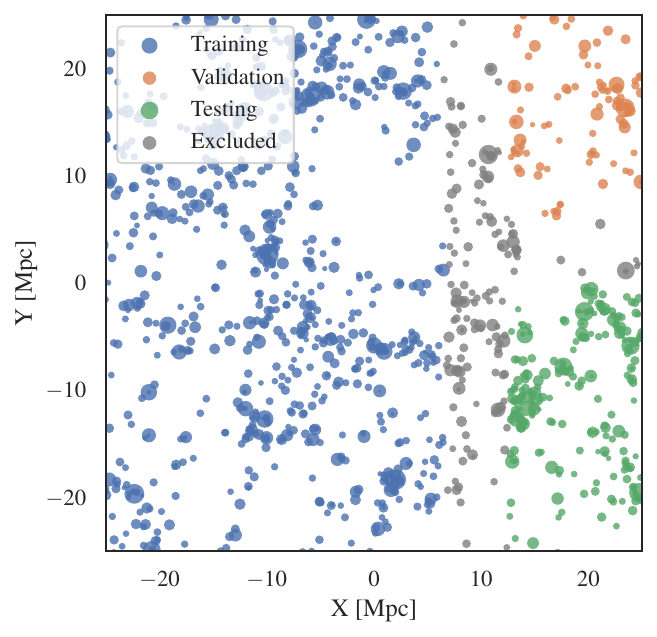}
    }
\subcaptionbox{Distributions of galaxy halo masses for data split.\label{fig:kde}}{
        \centering
        \includegraphics[width=80mm]{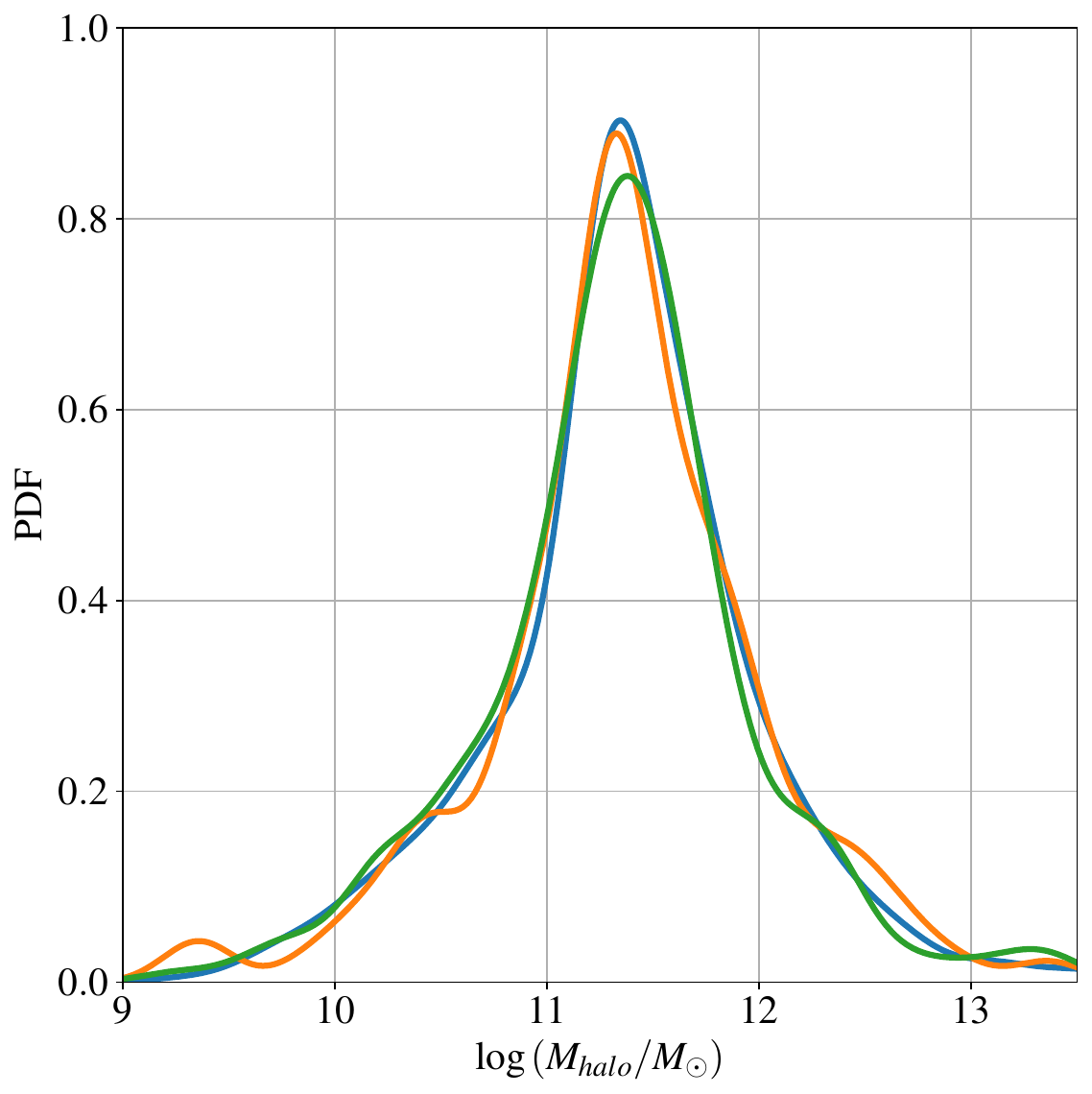}
        }
        \caption{Training, validation, and testing data split figures.}
    \label{fig:datasplit}
\end{figure*}
As is typical in ML problems, the available data are divided into training, validation and testing splits. In many cases, this is sampled randomly to reduce biases, but such a division is not appropriate for spatially correlated data such as graphs; GNNs use information from adjacent galaxies. Thus, a random split of the galaxy sample would cause information from the validation and testing sets to leak into the training set.

To completely separate these subsets of data, all galaxies (i.e., graph nodes) in the training, validation, and testing sets must be well-separated (i.e., they cannot be connected by graph edges). We set a constant linking length ($R_{max}$) of 3 Mpc during graph creation. We split the data into three contiguous sub-volumes, with a 6 Mpc partition in between each set to guarantee that the training, validation, and testing sets are independent. Figure \ref{fig:point-cloud} shows this data split in a two-dimensional projection. The resulting split leaves 1,011 galaxies for training, 112 galaxies for validation and 259 galaxies reserved for testing.

Due to our data set split, there is an increased risk that different subsets probe different cosmic environments. In other words, it is possible that the validation (or testing) sets comprise galaxies in a non-representative region of the Universe. This risk is exacerbated by the relatively small cosmic volume in the TNG50 simulation. To mitigate this risk, we check for differences between the halo mass distributions for the three data subsets. In Figure~\ref{fig:kde}, we show that the $M_{halo}$ distributions for each data set are quite similar. Thus, we proceed with this split and discuss the potential issue further in Section~\ref{sec:discussion}.

\subsection{Evaluation Metrics} \label{sec:metrics}

All models are trained to minimize Mean Squared Error (MSE). The test sets are evaluated on MSE, along with Mean Absolute Error (MAE),  Root Mean Square Error (RMSE), linear correlation coefficient ($R^2$), Normalized Median Absolute Deviation (NMAD), bias, and outlier fraction ($f_{outlier}$). 

NMAD is an outlier-insensitive metric for the prediction error (or scatter), normalized so that the NMAD of a Gaussian distribution is equal to a standard deviation.
\begin{equation}
    {\rm NMAD} \approx 1.4826\times \text{Median}\left(\left|y - \hat y -\text{Median}\left(y - \hat y\right)\right|\right).
\end{equation}

Bias is defined as the mean offset, and describes whether the parameter is generally overestimated or underestimated. 
\begin{equation}
    {\rm bias} = \sum_{i=1}^{N} (\hat y_i - y_i)
\end{equation}

The Outlier Fraction is the ratio of predictions with residuals $> 3 \times $ NMAD. This metric determines the rate of catastrophic outliers.
\begin{equation}
    f_{outlier} = \frac{1}{N} \cdot  \big|\big\{y_i : \left|\hat y - y\right| > 3 * \text{NMAD}\big\}\big|
\end{equation}

\subsection{Baseline Models} \label{sec:baseline-models}

\begin{table}[t]
\caption{Summary of TNG50 features used in our baseline models.\label{tab:baseline-features}}
\vskip 0.15in
\begin{center}
\begin{small}
\begin{sc}
\begin{tabular}{lccccr}
\toprule
Baseline Model  & $M_\star$ & $R_{Petro}$ & $S$  &  $A$ & $\Delta_G$ \\
\midrule
Simple  & $\surd$ & & &  \\
Morphological  & $\surd$ & $\surd$ & $\surd$ & $\surd$ &  \\
Overdensity & $\surd$ & & & & $\surd$  \\
Combined & $\surd$ & $\surd$ & $\surd$ & $\surd$ & $\surd$ \\
\bottomrule
\end{tabular}
\end{sc}
\end{small}
\end{center}
\vskip -0.1in
\end{table}

We train several baseline models to facilitate comparisons with deep neural networks; these baseline are not meant to achieve the best possible performance. Instead, they predict halo mass from commonly used galaxy features, such as stellar mass ($M_\star$), morphological parameters ($R_{Petro}, A, S$), and overdensity ($\Delta_G$). For each baseline model, we train a random forest with 100 estimators using the input features described below. Table~\ref{tab:baseline-features} summarizes how different baseline models employ the features described in Section~\ref{sec:data}.

\subsubsection{Simple Baseline Model} \label{sec:simple-baseline-model}
The simple baseline model is trained using stellar mass as the only feature. The galaxy-halo connection is often expressed as a one-to-one map between galaxy $M_\star$ and $M_{halo}$, so we expect that our simple baseline model should have some predictive power. However, there are secondary correlations between galaxy and dark matter halo properties \citep{2018ARA&A..56..435W}, and the simple baseline model will fail to capture those dependencies.

\subsubsection{Morphological Baseline Model} \label{sec:morphological-baseline-model}
The morphological baseline model is trained using several galaxy morphology features in conjunction with stellar mass. %
We use $k=5$-fold cross-validation to evaluate which morphological features are most critical to incorporate in our baseline model. Although we consider random forest models with the many morphological features described in Section~\ref{sec:morphological-features}, we find no substantial improvement after including Petrosian radius, Smoothness, and Asymmetry. Therefore, we elect to use those three morphological parameters in addition to stellar mass for our morphological baseline model.

\subsubsection{Overdensity Baseline Model}
There is evidence that galaxy overdensity can tighten the scatter in the SMHMR \citep[e.g.,][]{Blanton+2006}. Our overdensity model is a random forest that predicts $M_{halo}$ from $M_\star$ and $\Delta_G$.

\subsubsection{Combined Baseline Model}
We construct a combined baseline model using all features described above: $M_\star$, $R_{Petro}$, $A$, $S$, and $\Delta_G$. Any improvement in prediction for this model, relative to the other baseline models, can be interpreted as evidence that the galaxy morphology and large-scale environment contribute distinct information for estimating the halo mass.

\subsection{CNN}\label{sec:cnn}

\subsubsection{Model Architecture}

The backbone of the CNN is a ResNet18 \citep{DBLP:journals/corr/HeZRS15} pretrained on ImageNet data \cite{NEURIPS2019_9015}, with the final prediction layer replaced with 100 output features. The CNN output is concatenated with the galaxy's stellar mass, which is input into a 3 layer network to output the $M_{halo}$ estimate. All layers are trainable.

This CNN is trained for 1000 epochs with the AdamW optimizer \citep{kingma2014adam,loshchilov2019decoupled} at a learning rate of $\gamma=5\times10{-4}$, a weight decay of $\lambda=1\times10^{-3}$, and a batch size of 64.

\subsubsection{Image Data}

We apply point-wise Gaussian noise necessary to model sky background noise, with $\sigma$ values of $1/15$, $1/19$, and $1/25$ $e^- s^{-1}\text{pixel}^{-1}$ for the \textit{i}, \textit{r}, and \textit{g} band images. 
To reduce overfitting of the model, multiple simple augmentation techniques are used during model training:  horizontal flip, vertical flip, and rotation up to $90^\circ$. The image is padded by 5 pixels in every direction and randomly cropped, resulting in a random jitter. While a rigorous ablation study on data augmentation choices is outside the scope of our work, we ran simple validation experiments to determine that these augmentation techniques are helpful for efficiently training our model and reducing overfitting. We finally rescale the pixel values to take on a mean of 0 and standard deviation of 1 in each channel based on our (post-processed) image statistics; the same rescaling is applied to the training, validation, and testing data.

\subsection{GNN} \label{sec:gnn}

\subsubsection{Graph Construction}

Graphs are well-suited for modeling galaxies in cosmological volumes. Each node in the graph represents a galaxy, with stellar mass as the only node feature. We create edges between two galaxies if they are separated by less than the $R_{max} = 3$ used in the overdensity estimate; this is defined as the linking length of the graph. The Euclidean distance between connected nodes is used as the only edge feature.

After splitting the data into training, validation, and testing sets (see Section~\ref{sec:split}), we batch the training data set into 24 clusters using the METIS algorithm to reduce memory usage \citep[see, e.g., the \texttt{ClusterLoader} class in Pytorch-geometric][]{DBLP:journals/corr/abs-1905-07953,pyg}. 

\subsubsection{Model Architecture}

We use a GNN that can pass messages between edges from neighboring nodes, which enables node and edge states to be updated. This architecture permits the GNN to process information from each pair of neighboring nodes as well as the edge. A pooling layer then aggregates all of the information back to each node in order to simultaneously make predictions for every node in the batch.

Our GNN is based on the architecture described in \citet{wu2023learning}. Our GNN uses 4 parallel networks of fully-connected layers with 16 latent channels and 16 hidden channels, followed by \texttt{max} pooling. Each node's output is then concatenated with the original node feature (stellar mass), which is fed into a final three-layer neural network to predict $M_{halo}$.

We train the GNN for 500 epochs using the AdamW optimizer with a learning rate of $\gamma=1\times 10^{-3}$ and weight decay $\lambda = 1 \times 10^{-4}$. The entire validation and testing sets can each fit into a single batch, so we make predictions using the entire galaxy subgraph.

\subsection{CNN + GNN} \label{sec:cnn+gnn}

One final model combines the learnable parameters from both the CNN and the GNN. We construct the model using the (initially) frozen pretrained GNN backbone as the GNN component for this combined network. We also initialize the ImageNet-pretrained ResNet18 model for the CNN component, and attach a linear layer with 64 neurons. The GNN and CNN outputs are then concatenated with the stellar mass, which are passed through the same set of final linear layers used for the GNN.\footnote{The CNN outputs are not passed in as graph node features, as this was too computationally prohibitive for our experiments. Such a model may permit such a model to learn ``interactions'' between neighboring galaxies' image features.}

We batch examples using the same clustering algorithm that was used for the GNN model, which requires collating the same image examples for the CNN portion for model. This combined CNN+GNN is trained for 500 epochs using the AdamW optimizer with a learning rate of $\gamma=5\times 10^{-4}$ and $\lambda=1\times10^{-3}$, but because the CNN and GNN are initialized to pretrained weights, we use an early stopping criterion to refine the optimization procedure. If the validation MSE loss rises 0.005 above the minimum validation loss five times without finding a new minimum validation loss, then we re-enable learning on (i.e., unfreeze) the GNN backbone and reduce the learning rate by a factor of 5. If a new minimum validation loss is achieved, then the counter is reset to 0, and we proceed with training just the CNN. Unfreezing the GNN and  reducing the learning rate only occurs once; afterwards, we continue training for the remainder of the 500 epochs.

\section{Results}

\begin{figure*}[t]
    \subcaptionbox{Simple baseline}{
    \centering
    \includegraphics[width=40mm]{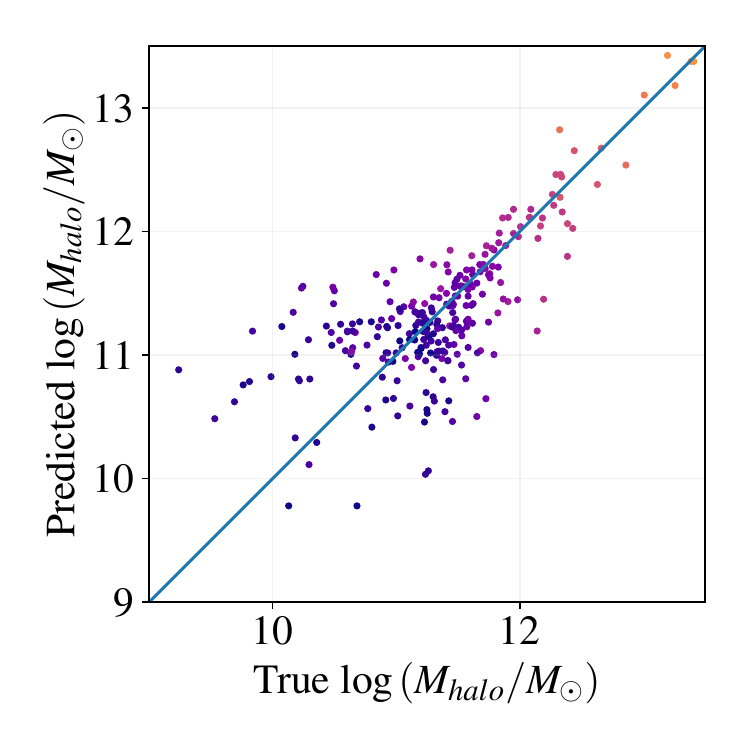}
    }
    \subcaptionbox{Morphological baseline}{
    \centering
    \includegraphics[width=40mm]{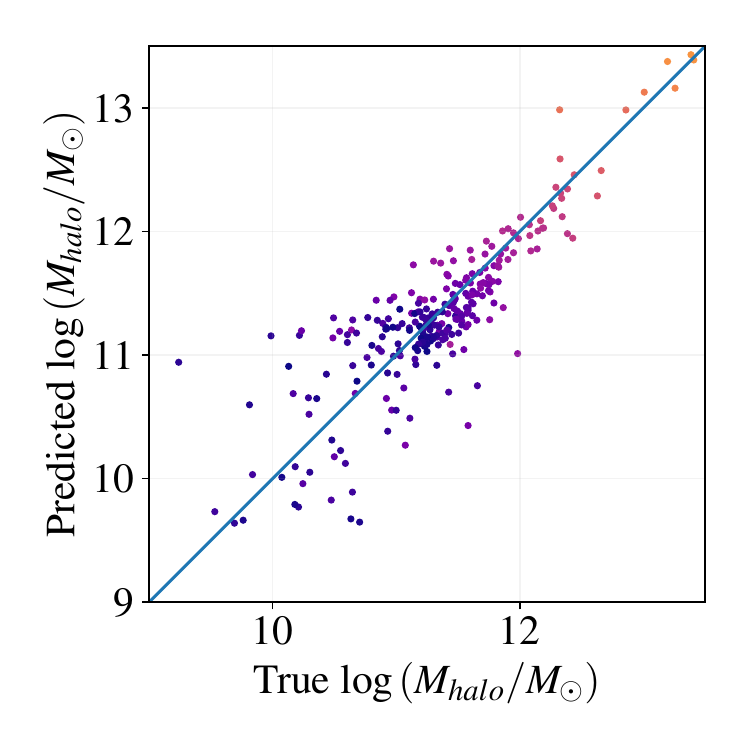}
    }
    \subcaptionbox{Overdensity baseline \label{overdensity_baseline}}{
    \centering
    \includegraphics[width=40mm]{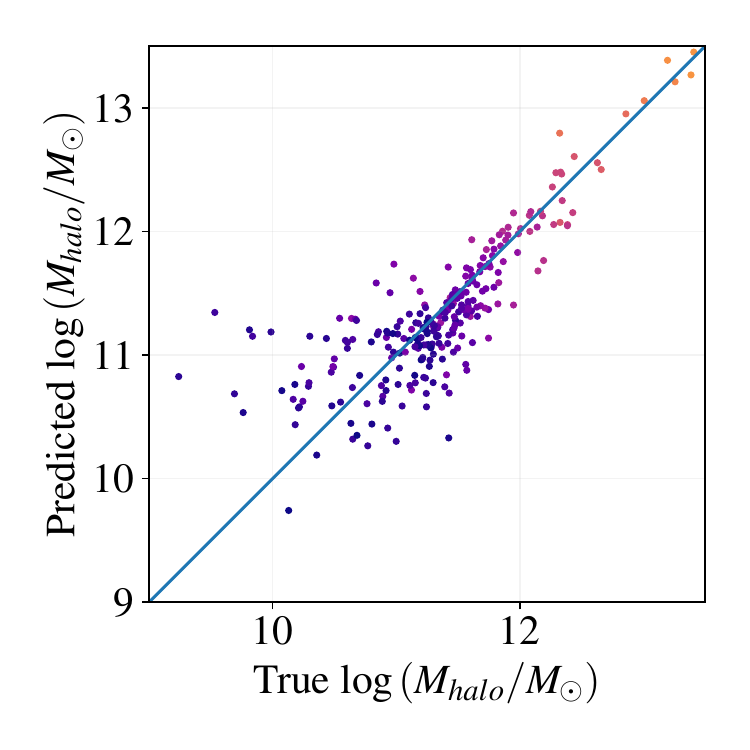}
    }
    \subcaptionbox{Combined baseline}{
    \centering
    \includegraphics[width=40mm]{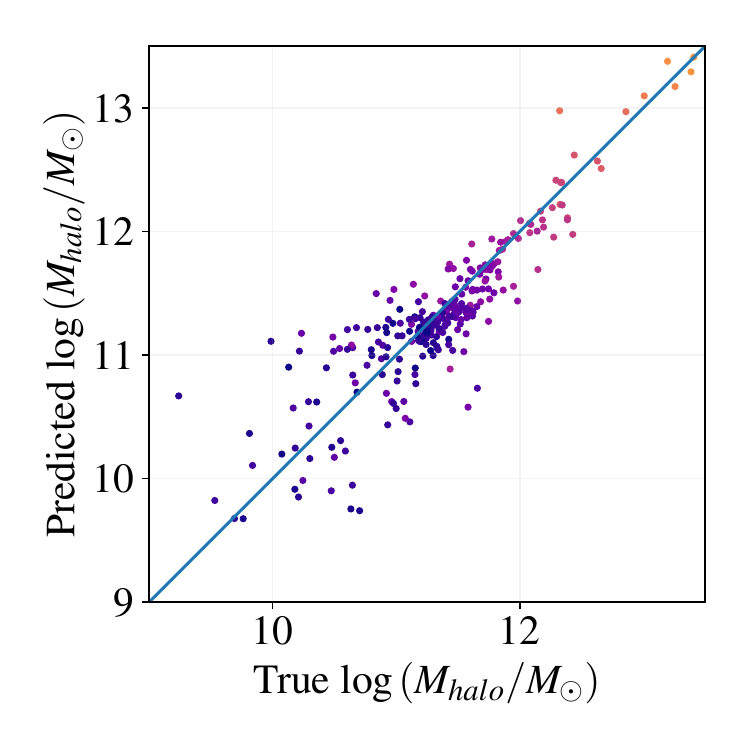}
    }
    \\
    \subcaptionbox{CNN\label{cnn_model}}{
    \centering
    \includegraphics[width=40mm]{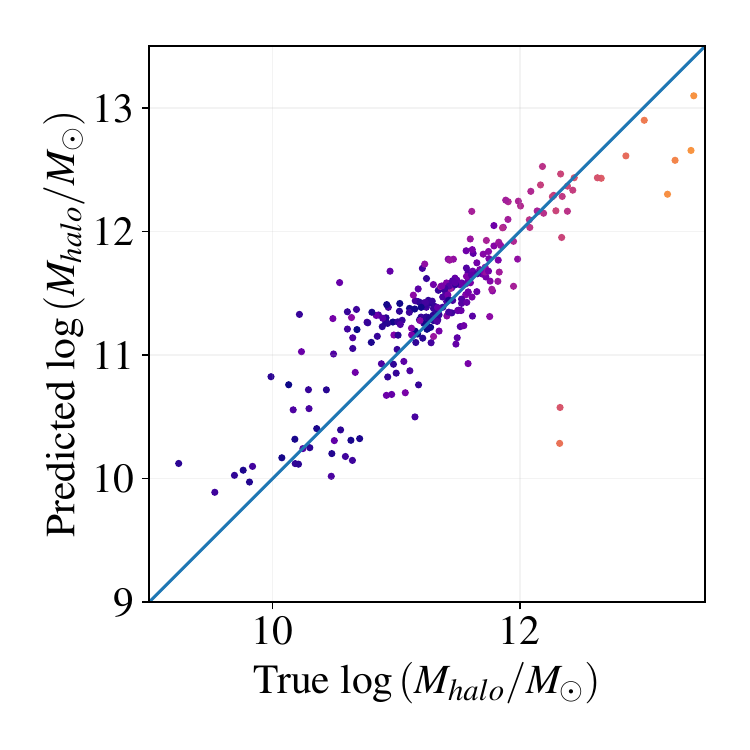}
    }
    \subcaptionbox{GNN}{
    \centering
    \includegraphics[width=40mm]{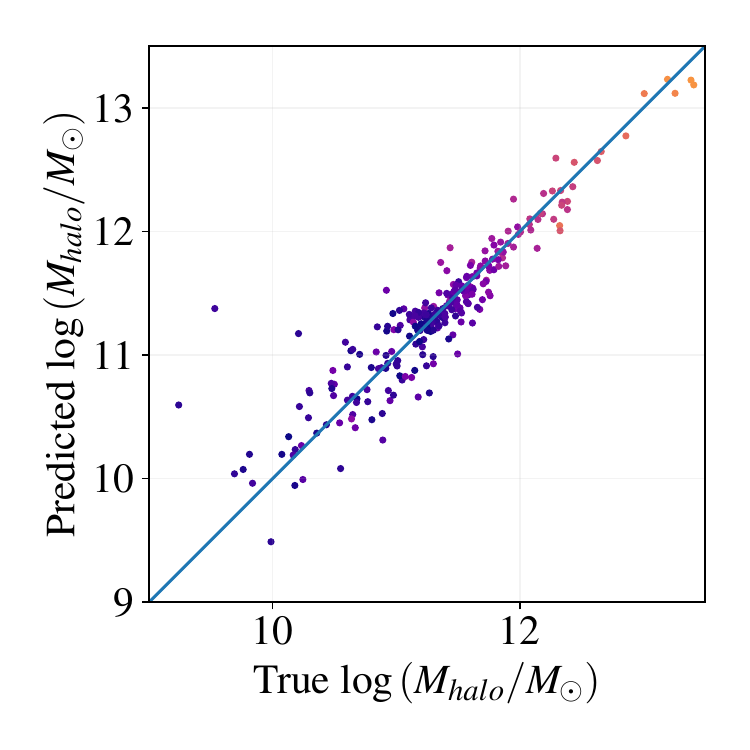}
    }
    \subcaptionbox{CNN+GNN\label{cnn+gnn_model}}{
    \centering
    \includegraphics[width=40mm]{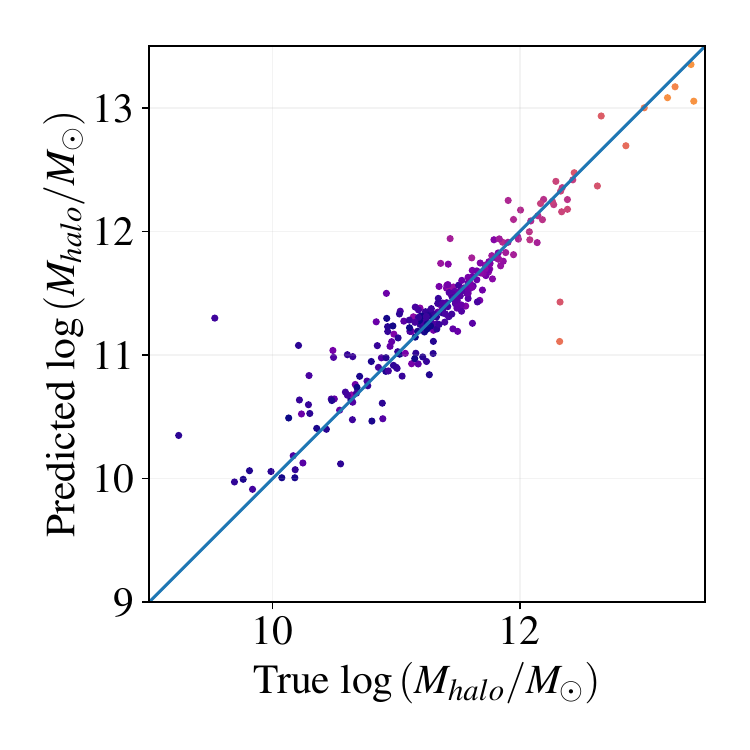}
    }
    \subcaptionbox{Stellar mass color scale}{
    \begin{minipage}[t]{40mm}
        \centering
        \includegraphics[width=17mm]{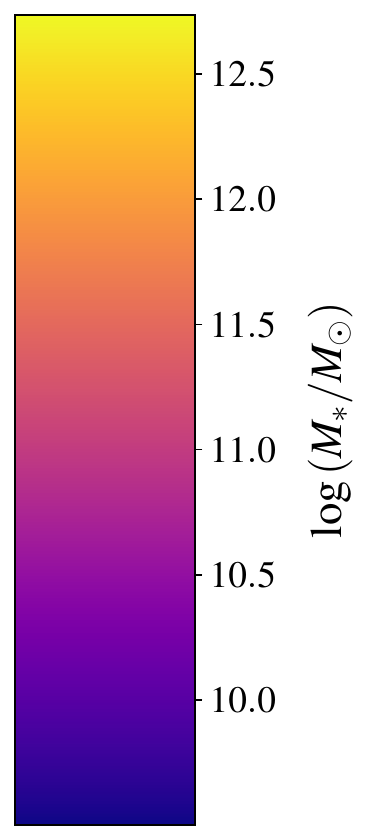}
    \end{minipage}
    }
    \caption{True vs. predicted $M_{halo}$ plots for all models, with color scale for $M_\star$. \label{fig:scatterplot-results}}
\end{figure*}

Our test set results are summarized in Figure~\ref{fig:scatterplot-results} and Table~\ref{tab:results}. The figure shows predicted versus true halo masses (colored by the stellar mass) for all models. The table compares the evaluation metrics (described in Section~\ref{sec:metrics}) for all models. 
Additional details about the neural network training and validation loss curves are presented in Appendix~\ref{appendix}.

\subsection{Baseline Models}

The simple baseline---a Random Forest using only stellar mass---performed the worst ($\text{RMSE}=0.455$), as expected. %
The morphological baseline model performs significantly better, with $\text{RMSE}=0.345$. There is clearly information encoded in these morphological features that cannot be captured by stellar mass alone. %
The overdensity baseline model performs better than the simple baseline ($\text{RMSE}=0.374$), but worse than the morphological baseline model. From Figure~\ref{overdensity_baseline}, we can see that it overpredicts the halo mass for most lower-mass systems, i.e., at true $\log(M_{halo}/M_\odot) < 10.5$. %
Finally, including both overdensity and the morphological features in the combined baseline model further reduces losses ($\text{RMSE}=0.310$). 

\subsection{CNN}

Surprisingly, we find that the CNN results in a poor RMSE (0.359) compared to the morphological baseline model (0.345), and suffers from a very high outlier fraction ($f_{outlier} = 6.18\%$). These outliers are visually apparent in Figure~\ref{cnn_model}, wherein the CNN fails to predict $M_{halo}$ for a few galaxies in the test set (and fails spectacularly for two galaxies in particular).
However, the MAE and NMAD is lower for the CNN than than for the morphological baseline model, suggesting that most predictions are fairly accurate aside from a few outliers.

\subsection{GNN}

The GNN predicts with much higher accuracy than the overdensity baseline, even though ResNet18 has hundreds of times more trainable parameters than the GNN. Relative to the previous models, the GNN has  few outliers and a low bias, indicating that the data it received does not differ largely from the training set.

\subsection{CNN + GNN}

The combined CNN+GNN model performs better than either the CNN and GNN separately, although the RMSE and $R^2$ values are very similar to those from the GNN. Again, we see the same two strong outliers (Figure~\ref{cnn+gnn_model}) as we did in the CNN-only model. However, the CNN+GNN still manages to achieve the best $f_{outlier}$, indicating that most of its predictions are very accurate. This is also supported by its low MAE and NMAD.

\begin{table*}[t]
\caption{Test results from baseline and neural network models as characterized by the evaluation metrics described in Section~\ref{sec:metrics}. The best metrics are underlined.}
\label{tab:results}
\vskip 0.15in
\begin{center}
\begin{small}
\begin{sc}
\begin{tabular}{lccccccc}
\toprule
Model Name      & MSE   & MAE   & RMSE  & $R^2$ & Bias     & NMAD  & $f_{outlier}$ \\
\midrule
Simple Baseline   & 0.207 & 0.328 & 0.455 & 0.524 & $+$0.009 & 0.313 & 2.27\%   \\
Morphological Baseline   & 0.119 & 0.246 & 0.345 & 0.726 & $-$0.023 & 0.234 & 2.43\%   \\
Overdensity Basline    & 0.140 & 0.260 & 0.374 & 0.680 & $-$0.008 & 0.261 & 2.09\%   \\
Combined Baseline    & 0.096 & 0.220 & 0.310 & 0.779 & $-$0.034 & 0.220 & 2.26\%   \\
\midrule
CNN       & 0.129 & 0.238 & 0.359 & 0.705 & $+$0.062 & 0.218 & 6.18\%   \\
GNN       & 0.062 & 0.158 & 0.248 & 0.858 & \underline{$+$0.007} & 0.146 & 1.78\%   \\
CNN + GNN & \underline{0.062} & \underline{0.144} & \underline{0.248} & \underline{0.859} & $+$0.020 & \underline{0.123} & \underline{1.77\%}   \\
\bottomrule
\end{tabular}
\end{sc}
\end{small}
\end{center}
\vskip -0.1in
\end{table*}

\section{Discussion} \label{sec:discussion}

\subsection{Deep neural networks surpass baseline models}
Results from our baseline models demonstrate that galaxy morphology and environment are valuable for improving $M_{halo}$ predictions for TNG50 galaxies. Moreover, these features can be combined to achieve even better results. The baseline models use ``handcrafted'' features that are commonly used in astronomy \citep[e.g.,][]{2003ApJS..147....1C}, but these features may not summarize all the information in galaxy images or environments that are relevant for constraining $M_{halo}$. The successes of deep learning offers a hint that CNNs and GNNs may be able to extract more useful information directly from images and point clouds. Here, we interpret our results from more sophisticated CNNs and GNNs.

We find that the neural network models generally perform better than the baseline models (Table~\ref{tab:results}), confirming our previous hypothesis. This is despite the fact that highly overparameterized neural networks can struggle to learn from small data sets (see Section~\ref{sec:cosmic-variance} for more discussion). Nonetheless, the CNN still performs better than the morphological baseline model in terms of outlier-insensitive metrics such as the MAE and NMAD. The GNN learns far more environmental information than is encoded in the simple overdensity parameter $\Delta_G$, leading to far lower prediction errors.\footnote{We also note that the overdensity baseline model performs very poorly for low-mass galaxies, which may indicate challenges with predicting $M_{halo}$ for satellite galaxies residing in high-density environments.} The CNN+GNN model achieves the best performance metrics across the board---but it could probably perform even better if not for the CNN's catastrophic failures.

\subsection{Successes and failures of neural networks} \label{sec:discuss-nns}

The CNN primarily fails in two different ways. First, the halo masses of high-mass ($M_\star > 10^{12.5} M_\odot$) galaxies are consistently underpredicted. The latter failure mode can be explained by small TNG50 volume, which inhibits the CNN---with its $\mathcal O(10^7)$ trainable parameters---from adequately learning how to handle these rare massive galaxies. Second, there are two galaxy samples in the test set that have dramatically underpredicted halo masses (see, e.g., Fig~2e). We have visually inspected these galaxy images, and found that one of them is a pair of interacting galaxies.
As we note in Appendix~\ref{appendix} (see Figure~\ref{fig:losses}), the validation loss achieves a minimum value after only 116 epochs of training, which suggests that the CNN is undertrained. The validation loss of $\text{MSE}=0.0894$ is much better than test loss $\text{MSE}=0.129$; thus we surmise that the CNN did not fully converge, and would benefit from more training examples. 

It is interesting that the GNN demonstrates such strong performance in predicting $M_{halo}$. One interpretation of this result is that the large-scale environment is more informative for predicting $M_{halo}$ than the detailed galaxy appearances. However, the morphological baseline model \textit{does} outperform the overdensity baseline model, which may imply that our overdensity parameter $\Delta_G$ is too simplistic to describe the overall environment. Indeed, \citet{wu2024galaxyhalo} find that GNNs are better suited than overdensity for describing environmental dependence on the relationship between galaxies and their dark matter halos, which reinforces the idea that GNNs are better equipped to extract environmental information from galaxy point cloud data.

\subsection{The challenges of learning from small data sets} \label{sec:cosmic-variance}

We have mentioned that our data split reduces the galaxy data set to only a about a thousand examples for training, and a few hundred examples each for validation and testing. The limited data sets raise several challenges. First, neural networks typically benefit from more training data, so it is likely that our deep learning models have the capacity for further optimization. Second, the small subsets of data for validation or testing can increase the variance in prediction results; it is conceivable that a nested cross-validation can help ameliorate this issue. Third, due to our choice to separate galaxies into contiguous subvolumes, the validation or testing data sets can exhibit different galaxy properties on average than the ones seen in the training data set.

We anticipate that our neural networks would perform even better or larger data volumes. However, increasing the simulation box size often comes at the cost of lowering the resolution, which can have an adverse effect on simulating realistic galaxy appearances at all. 

\subsection{Application to real data and domain adaptation} \label{sec:discuss-caveats}

Our successful experiments indicate that we may be able to better predict galaxy properties by leveraging galaxy imaging and their cosmic surroundings. Can we immediately apply this to real observations? Unfortunately, the answer is likely not. Our models have learned the very specific characteristics of simulated data from TNG50, which differs from the real Universe in myriad ways. In general, this problem is known as ``domain shift'' or ``data set shift'' and it applies to any kind of shift in the high-dimensional inputs to deep learning models. Not only does domain shift prevent us from successfully applying a model trained on TNG50 to the real Universe, but it also prevents us from applying this model trained on TNG50 to another simulation. There exist proposed solutions for mitigating this effect \citep[``domain adaptation,'' see e.g.,][]{Csurka2017}, which have been applied to astronomical data \citep{Ciprijanovic+2023}, but the problem is far from solved. Therefore, we caution against naively using these ML models to predict $M_{halo}$ for real galaxies.

\section{Conclusions}

We have estimated the dark matter halo mass, $M_{halo}$, using galaxy data from the Illustris TNG50-1 hydrodynamic simulations. We split our sample of well-resolved galaxies with reliable morphological measurements into spatially separated regions for the training ($N=1011$), validation ($N=112$), and testing ($N=259$) data subsets. All results are presented using metrics from the testing data set. 

We first train baseline random forest models that use stellar mass ($M_\star$), morphological features (Petrosian radius, asymmetry, and smoothness), and galaxy overdensity ($\Delta_G$) over 3 Mpc scales. From our baseline model investigation, we find that galaxy morphology and overdensity are useful features for accurately estimating $M_{halo}$. Moreover, the combined set of features produces an even more performant model (Table~\ref{tab:results}).

After we confirm that morphology and overdensity are important for predicting dark matter halo mass, we use synthetic galaxy image cutouts and galaxy $3d$ point cloud as inputs to neural network models. In other words, we eschew summary statistics in favor of directly learning from the pixel-level and point cloud data (in addition to the stellar mass). Our conclusions for the neural network models are listed below:
\begin{enumerate}
    \item While a deep convolutional neural network (CNN) can achieve strong performance on the training and validation set ($\text{RMSE}=0.299$), its performance on the test set suggests that it has been overfit to the small data set ($\text{RMSE}=0.359$, which is worse than the morphological baseline model result---$\text{RMSE}=0.345$). This interpretation is also supported by its extremely poor performance on two dramatic outliers in the test set (see Figure~\ref{cnn_model}) and its large outlier fraction ($f_{outlier} = 6.18\%$). Nevertheless, the CNN outperforms the morphological baseline well in terms of outlier-insensitive metrics like mean absolute error ($\text{MAE}=0.238$) and normalized median absolute deviation ($\text{NMAD}=0.218$).
    \item The graph neural network (GNN), trained on cosmic graphs connected by a 3 Mpc linking length, achieves very low prediction error ($\text{RMSE}=0.248$) relative to the overdensity baseline model ($\text{RMSE}=0.374$) and combined baseline model ($\text{RMSE}=0.310$), which suggests that galaxy environment is particularly important for constraining the halo mass.
    \item We have trained a novel CNN+GNN joint model that achieves the best performance overall. Because the CNN component of the model suffers from the overfitting issue described above (see also Figure~\ref{cnn+gnn_model}), the CNN+GNN model is comparable to the GNN model in terms of RMSE and $R^2$; however, its outlier-insensitive metrics ($\text{MAE}=0.144$, $\text{NMAD}=0.123$) are far superior to any other models' performance.
\end{enumerate} 

Our results demonstrate that a CNN+GNN model is capable of jointly extracting detailed information from galaxy appearances and large-scale environments. These results are promising for future data-driven approaches to predicting dark matter properties from galaxies, particularly if they can be trained on much larger galaxy samples spanning larger cosmic volumes (see Section~\ref{sec:cosmic-variance}). However, we also caution against applying these models trained on simulation data to real observations without first accounting for domain shift (Section~\ref{sec:discuss-caveats}).

\bibliography{paper}
\bibliographystyle{icml2024}

\newpage
\appendix
\onecolumn
\section{Training and validation loss curves} \label{appendix}

\begin{figure}[h]
    \centering
    \subcaptionbox{CNN: minimum validation loss during epoch 116. \label{fig:loss-cnn}}{
    \centering
    \includegraphics[width=120mm]{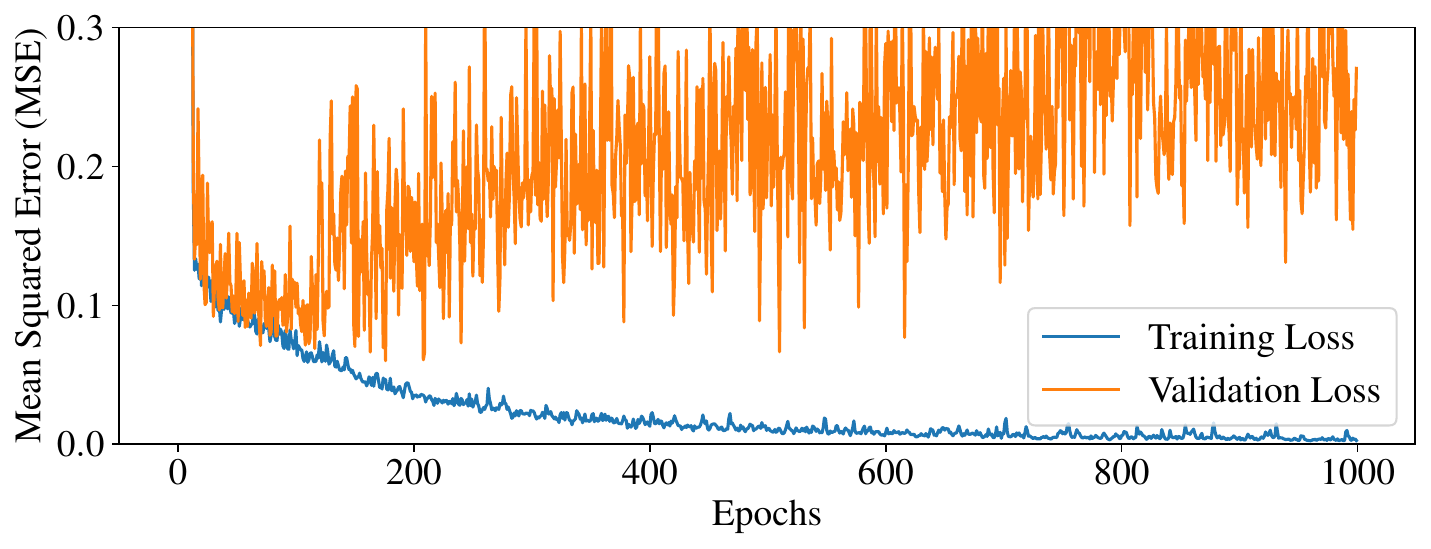}
    }
    \subcaptionbox{GNN: minimum validation loss during epoch 416.}{
    \centering
    \includegraphics[width=120mm]{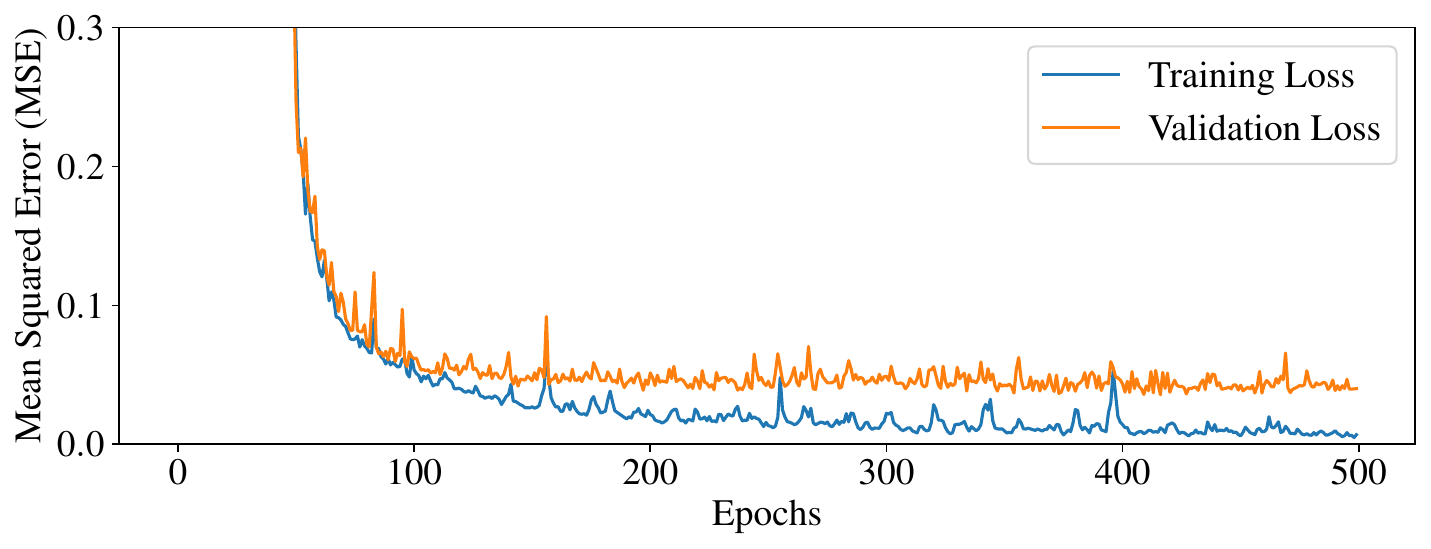}
    }
    \subcaptionbox{CNN+GNN: After epoch 144, we unfroze the GNN component and lowered the CNN learning rate (see description of optimization procedure in Section~\ref{sec:cnn+gnn}). Minimum validation loss during epoch 466.}{
    \centering
    \includegraphics[width=120mm]{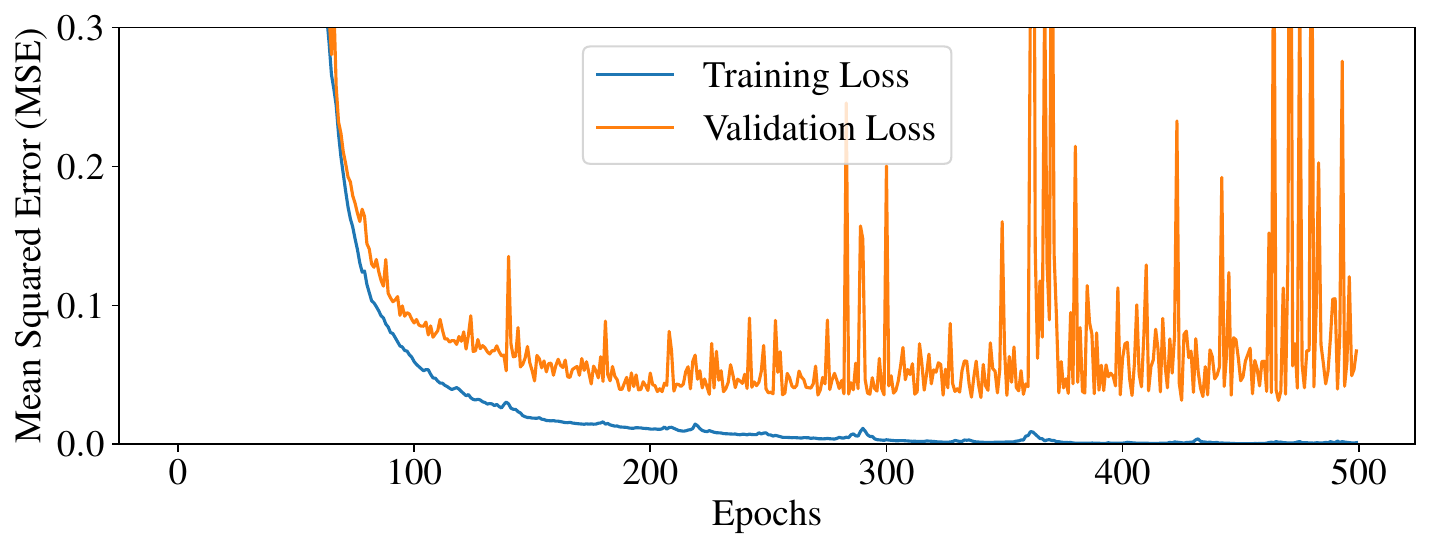}
    }
    \caption{Training and validation loss curves for our deep learning models. The subfigure caption indicates which epoch minimized the validation loss; that checkpointed model is then used to make final test set predictions. \label{fig:losses}}
\end{figure}

In Figure~\ref{fig:losses}, we show the training and validation loss curves from optimizing the CNN, GNN, and CNN+GNN models via the training procedures described in Section~\ref{sec:experiments}. To make predictions, we save the model checkpoints that achieve lowest validation losses, but here we show the entire loss curves (out to 500 epochs) in order to gauge the level of over- or under-fitting.

Figure~\ref{fig:loss-cnn} shows that the CNN validation loss curve begins to increase shortly after the minimum validation epoch (116) while the training loss continues to decrease. The validation loss is quite noisy, which suggests that it may not be a reliable indicator of model convergence. Thus, the CNN's modest results are unsurprising (e.g., Section~\ref{sec:discuss-nns}).

The GNN and CNN+GNN loss curves (in panels b and c) seem to indicate that there is a substantial gap between training and validation losses. However, when optimizing GNNs, it is common to find that the training and validation loss curves plateau rather than diverge---which would signify a problem with overfitting. This is likely because GNNs have far fewer trainable parameters than CNNs, and are therefore less susceptible to overfitting.

\end{document}